\documentclass[
article,
superscriptaddress,
amsmath,
amssymb,
aps,
pra, 
longbibliography,
floatfix
]{revtex4-1}

\usepackage{setspace}
\usepackage{float}
\usepackage{graphicx}
\usepackage{xcolor}
\usepackage{bm}
\usepackage[mathlines]{lineno}
\usepackage{tabularx}

\usepackage{comment}
\usepackage{enumitem}

\usepackage{textcomp,mathcomp}

\usepackage{color}
\usepackage{dcolumn}

\usepackage[english]{babel}
\usepackage[latin1]{inputenc}
\usepackage[bottom=2.5cm,top=2.5cm, left=2cm, right=2cm]{geometry}
\usepackage{siunitx}
\usepackage{xr}
\makeatletter
\newcommand*{\addFileDependency}[1]{
  \typeout{(#1)}
  \@addtofilelist{#1}
  \IfFileExists{#1}{}{\typeout{No file #1.}}
}
\makeatother

\newcommand*{\myexternaldocument}[1]{
    \externaldocument{#1}
    \addFileDependency{#1.tex}
    \addFileDependency{#1.aux}
}

\myexternaldocument{Supplementary_Materials}

\usepackage[unicode=true, bookmarks=true, bookmarksnumbered=false, bookmarksopen=false, breaklinks=false, pdfborder={0 0 1}, backref=false, colorlinks=false]{hyperref}

\usepackage{xcolor}

\usepackage{soul}

\begin{document}
\setstretch{2.75}
\title{Nanoscale electrostatic control in ferroelectric thin films through lattice chemistry}

\author{Ipek Efe}
\email[]{ipek.efe@mat.ethz.ch}
\affiliation{Department of Materials, ETH Zurich, CH-8093 Zurich, Switzerland}
\author{Alexander Vogel}
\affiliation{Electron Microscopy Center, Empa, CH-8600 Dubendorf, Switzerland}
\affiliation{Swiss Nanoscience Institute, University of Basel, CH-4056 Basel, Switzerland}
\author{William S. Huxter}
\affiliation{Department of Physics, ETH Zurich, CH-8093 Zurich, Switzerland}
\affiliation{Quantum Center, ETH Zurich, CH-8093 Zurich, Switzerland}
\author{Elzbieta Gradauskaite}
\affiliation{Department of Materials, ETH Zurich, CH-8093 Zurich, Switzerland}\author{Iaroslav Gaponenko}
\affiliation{Department of Quantum Matter Physics, University of Geneva, 1211 Geneva, Switzerland}
\author{Patrycja Paruch}
\affiliation{Department of Quantum Matter Physics, University of Geneva, 1211 Geneva, Switzerland}
\author{Christian L. Degen}
\affiliation{Department of Physics, ETH Zurich, CH-8093 Zurich, Switzerland}
\affiliation{Quantum Center, ETH Zurich, CH-8093 Zurich, Switzerland}
\author{Marta D. Rossell}
\affiliation{Electron Microscopy Center, Empa, CH-8600 Dubendorf, Switzerland}
\author{Manfred Fiebig}
\affiliation{Department of Materials, ETH Zurich, CH-8093 Zurich, Switzerland}
\author{Morgan Trassin}
\email[]{morgan.trassin@mat.ethz.ch}
\affiliation{Department of Materials, ETH Zurich, CH-8093 Zurich, Switzerland}

\date {\today}

\begin{abstract}

Nanoscale electrostatic control of oxide interfaces enables physical phenomena and exotic functionalities beyond the realm of the bulk material. In technologically-relevant ferroelectric thin films, the interface-mediated polarization control is usually exerted by engineering the depolarizing field. Here, in contrast, we introduce polarizing surfaces and lattice chemistry engineering as an alternative strategy. Specifically, we engineer the electric-dipole ordering in ferroelectric oxide heterostructures by exploiting the charged sheets of the layered Aurivillius model system. By tracking in-situ the formation of the Aurivillius charged Bi$_{2}$O$_{2}$ sheets, we reveal their polarizing effect leading to the characteristic Aurivillius out-of-plane antipolar ordering. Next, we use the polarizing Bi$_{2}$O$_{2}$ stacking as a versatile electrostatic environment to create new electric dipole configurations. We insert multiferroic BiFeO$_3$ into the Aurivillius framework to stabilize a ferrielectric-like non-collinear electric-dipole order in the final heterostructure while maintaining the antiferromagnetic order of BiFeO$_3$. We thus demonstrate that engineering the lattice chemistry stabilizes unconventional ferroic orderings at the nanoscale, a strategy that may be expanded beyond the realm of electrically ordered materials.

\end{abstract}

\maketitle


 Advances in the design of ferroelectric$|$dielectric superlattices with well-defined electrostatic and elastic boundary conditions facilitated the creation of topological electric-dipole orderings with technological significance \cite{Topological_review_das_new_2020}. For instance, the observation of polar skyrmions  \cite{das_observation_2019}, ferroelectric vortices \cite{VORTICES_yadav_observation_2016}, and merons \cite{wang_polar_2020} expanded the range of potential applications for epitaxially stacked ferroelectric materials \cite{FESO_noel_non-volatile_2020,marvin_muller_ferroelectric_2023}. Recently, the superlattice design has even enabled the generation of an antiferroelectric phase \cite{mundy_liberating_2022} as well as of functional properties such as negative capacitance \cite{NC_zubko_negative_2016} and emergent chirality \cite{shafer_emergent_2018}.
Typically, the superlattice approach towards electrostatic engineering aims at tuning the stability of the polarization in a ferroelectric constituent. This has been realized so far by balancing elastic boundary conditions and uncompensated-bound-charge-driven depolarizing-field contributions \cite{das_observation_2019} at the interface. However, the required balance between the elastic and electrostatic interactions can only be achieved for a limited number of materials, and the resulting compensated electric dipole ordering limits controllability by applying an external electric field. Thus, alternatives to such depolarizing-field engineering are highly desired and will expand the means for electrostatic control of oxide interfaces.

Exploiting naturally occurring modifications of the surface chemistry in oxide ferroelectric thin films as polarizing layers \cite{spaldin_layer_2021,efe_engineering_2024}, rather than interfacing the films with dielectrics to tune the depolarizing field \cite{das_observation_2019,VORTICES_yadav_observation_2016,wang_polar_2020}, is a promising alternative route towards control of electrostatic boundary conditions. For instance, an off-stoichiometric charged surface layer may form spontaneously in oxide materials \cite{Nives_strkalj_-situ_2020,xie_giant_2017,li_control_2018}. For ferroelectric materials, the resulting charge gradient has been shown to enhance the polarization \cite{Nives_strkalj_-situ_2020,xie_giant_2017} and orient the electric dipole formation \cite{li_control_2018}.
Hence, the integration of such charged layers with well-defined chemistry and charge density into ferroelectric heterostructures would provide a disruptive advance from interface-physics-based to chemical-engineering-based control \cite{efe_engineering_2024} of electric-dipole textures. As a particular advantage and in striking contrast with the depolarizing-field approach, polar configurations may be enhanced or created, thus extending the materials systems beyond the realm of ferroelectrics, such as dielectrics at large, magnetic, or superconducting systems. Controlling the formation of such charged layers to enable atomic-scale poling of electric dipoles, however, has remained challenging. Charged layers tend to nucleate spontaneously, preventing true control over the electrostatic configuration.

Here, we use the charged, polarizing building blocks of layered Aurivillius compounds to control the electric-dipole ordering in oxide superlattices at the nanoscale. We track the polarization dynamics during the growth of the model system Bi$_{5}$FeTi$_{3}$O$_{15}$ (BFTO) using optical in-situ second harmonic generation (ISHG). In a first step, we enable the controlled coverage of the surface with charged Bi$_{2}$O$_{2}$ sheets. This results in an abrupt reorientation of electric dipoles in the films and causes their characteristic antipolar ordering perpendicular to the Aurivillius layering. By incorporating additional functional constituents into the Aurivillius crystal structure, we expand the concept of lattice chemistry engineering for atomic-scale poling. Specifically, by integrating multiferroic BiFeO$_3$ (BFO) into the layered Aurivillius architecture, we obtain a transition from the classical out-of-plane antipolar ordering of the pure Aurivillius compounds towards a non-collinear ferrielectric-like electric-dipole ordering in the composite structure. Preservation of the characteristic antiferromagnetic order of BFO elegantly demonstrates our capability to merge the properties of the two parent compounds while achieving precise control over their electric-dipole configuration. Looking forward, inserting other functional building blocks, such as magnetically ordered or superconducting perovskites, into the layered framework will expand the use of lattice chemistry and polarizing layers for atomic-scale poling beyond the realm of ferroelectrics.


The intimate connection between the charged-layered structure and polarization in the Aurivillius compounds opens up possibilities for the design of exotic polarization states \cite{moore_charged_2022} and motivates our selection of such model system for tailoring electric-dipole ordering using lattice chemistry. We start our investigation with a close look at the Aurivillius BFTO unit cell. 
The two building blocks of the unit cell, i.e., the slab of four perovskite layers Bi$_3$FeTi$_3$O$_{13}$ and the Bi$_{2}$O$_{2}$ fluorite-like layer, are shown in Fig. 1a.
The polarization in BFTO emerges predominantly as a consequence of the cation displacements within the \textit{AB}O$_{3}$-type perovskite blocks \cite{Auri_ferroelectricity_1_newnham_structural_1971,Auri_ferroelectricity_2_withers_crystal_1991,Auri_ferroelectricity_3_hervoches_structural_2002,Auri_ferroelectricity_4_benedek_understanding_2015,Auri_ferroelectricity_5_birenbaum_potentially_2014}, where the Bi cations are placed at the \textit{A}-sites, while Ti and Fe cations occupy the \textit{B}-sites. 
In Aurivillius compounds, the high negative-charge density localized at the (2O)$^{4-}$ oxygen plane within the fluorite-like layers \cite{gradauskaite_defeating_2023,spaldin_layer_2021} compensates positive bound charges. This interplay dictates the polarization orientation of the perovskite units. Thus, in stark contrast with the classical perovskite ferroelectrics with their uniform electric-dipole orientation, the \textit{c}-axis-oriented Aurivillius structure exhibits out-of-plane antipolar ordering, as shown in Fig. 1a. In particular, in  Bi$_{n+1}$Fe$_{n-3}$Ti$_3$O$_{3n+3}$ thin films with even \textit{n}, the distribution of electric dipoles orientation splits evenly towards the top and bottom Bi$_{2}$O$_{2}$ layers and causes the suppression of the net out-of-plane polarization  \cite{campanini_buried_2019,moore_charged_2022,Auri_ferroelectricity_5_birenbaum_potentially_2014,sun_progress_2021, gradauskaite_nanoscale_2021}. The configuration for BFTO (\textit{n}=4), one of the most debated representatives of the Aurivillius compounds \cite{Auri_ferroelectricity_3_hervoches_structural_2002,kubel_x-ray_1992,Auri_ferroelectricity_5_birenbaum_potentially_2014,gradauskaite_robust_2020,li_control_2018, Keeney_in_plane_3_zhang_structural_2012}, is resolved by high-angle annular dark field scanning transmission electron microscopy (HAADF-STEM) in Fig. 1b. In particular, electric-dipole mapping reveals the drastic suppression of the net out-of-plane polarization component, fully consistent with reports of the net polarization cancellation at a larger scale based on macroscopic ferroelectric characterizations \cite{Keeney_in_plane_3_zhang_structural_2012,gradauskaite_robust_2020}.

To take advantage of such charged-layered architecture to engineer a nanoscale polarization state beyond the realm of existing Aurivillius ferroelectric phases, a precise understanding of the electric-dipole ordering dynamics during the epitaxial design is, however, an absolute prerequisite. Here we achieve such polarization evolution tracking during the thin-film deposition process by ISHG \cite{sarott_situ_2021,de_luca_nanoscale_2017}. Optical second harmonic generation (SHG) denotes the frequency doubling of light in a material. In the electric-dipole approximation of the light field SHG is allowed in systems with broken inversion symmetry, which makes this technique a very powerful probe of ferroelectric polarization \cite{nordlander_probing_2018,denev_probing_2011}. Simultaneous reflection high energy electron diffraction (RHEED) monitoring permits us to calibrate the ISHG intensity to the film thickness with sub-unit cell accuracy (see Methods).
%

The evolution of the RHEED signal during the pulsed-laser-deposition growth of BFTO on a lattice-matching (001)-oriented orthorhombic NdGaO$_{3}$ (NGO) substrate is shown in Fig. 2a (see Methods for growth conditions). The oscillating RHEED intensity indicates layer-by-layer growth of the BFTO \cite{gradauskaite_robust_2020}. Ex-situ X-ray-reflectivity (XRR) correlates a single RHEED oscillation to half a BFTO  unit cell ($\approx$ 2 nm). 
The thin-film deposition takes place below the ferroelectric transition temperature (T$_{\text{C}}$ $=$ 730 -- 750 \textdegree C $\gg$ T$_{\text{growth}}$  $=$ 700 \textdegree C) \cite{li_ferroelectric_2010}. This permits us to probe the emerging net polarization during the growth, which we achieve by recording the ISHG signal in two optical configurations to independently capture the net in-plane ({\textit{P}$\mathrm{_{net}}$}$\mathrm{^{IP}}$) and the net out-of-plane ({\textit{P}$\mathrm{_{net}}$}$\mathrm{^{OOP}}$) polarization components in our films (see Methods).
The evolution of the associated ISHG signals for BFTO is shown in  Fig. 2b. We first notice that the ISHG signal related to {\textit{P}$\mathrm{_{net}}$}$\mathrm{^{IP}}$ remains at the paraelectric background level throughout the entire growth, which points to {\textit{P}$\mathrm{_{net}}$}$\mathrm{^{IP}}$ = 0. This is surprising since BFTO is well established as an in-plane-polarized ferroelectric material \cite{keeney_persistence_2020,Keenay_in_planeP_1_faraz_study_2015,Keeney_in_plane_3_zhang_structural_2012,gradauskaite_robust_2020,sun_progress_2021,song_evolution_2018}. In this uniaxial ferroelectric material \cite{gradauskaite_robust_2020}, we suspect that a dense network of oppositely in-plane-polarized domains leads to the cancellation of {\textit{P}$\mathrm{_{net}}$}$\mathrm{^{IP}}$ and, hence, of the ISHG signal in this configuration \cite{denev_probing_2011,gradauskaite_defeating_2023}. We corroborate this by ex-situ piezoresponse force microscopy (PFM) (Fig. S1). 

In contrast, the ISHG signal from {\textit{P}$\mathrm{_{net}}$}$\mathrm{^{OOP}}$ yields a striking periodic, saw-tooth-like evolution.
The out-of-phase behavior of the RHEED and ISHG signals rules out surface contributions \cite{jahnke_shg_1999,vollmer_magnetization_2001} as the source of the ISHG oscillation. Furthermore, the pronounced saw-tooth-like ISHG response differs from the previously reported ISHG oscillations caused by structural symmetry breaking during the growth of non-polar oxide systems \cite{nordlander_inversion-symmetry_2021}. 
Therefore, our ISHG signal must have a non-trivial origin. We ascribe the ISHG oscillations to the periodic evolution of the out-of-plane polarization caused by the BFTO layer-by-layer growth dynamics.

The period of the ISHG oscillation is equal to the period of a RHEED oscillation and thus corresponds to the growth of half a unit cell of BFTO. 
In the light of the preceding discussion, it is reasonable to assume that for each half unit cell of BFTO, {\textit{P}$\mathrm{_{net}}$}$\mathrm{^{OOP}}$ builds up with the ongoing coverage by the perovskite blocks. The ISHG signal then undergoes an abrupt, saw-tooth-like cancellation with the deposition of the Bi$_2$O$_2$ sheet, which imposes the antipolar ordering in the completed half unit cell and cancels {\textit{P}$\mathrm{_{net}}$}$\mathrm{^{OOP}}$ (see Fig. 2a-b).
Ex-situ atomic force microscopy (AFM) of the thin-film morphology after growth termination at different stages of the oscillating RHEED and ISHG signals reveals that films whose growth was stopped either at an ISHG maximum or an ISHG minimum exhibit identical surface topographies (Fig. S2a--c). This is consistent with the complete, uniform capping of the Bi$_{2}$O$_{2}$ on top of the perovskite blocks, which leaves the topography unaffected but triggers the antipolar order in the films, as sketched in Fig. 2c. During the stage of the deposition with continuous ISHG signal increase, the ex-situ AFM scans show half-unit-cell-high features progressively covering the surface (Fig. S2). This observation corroborates that the perovskite blocks carry a non-zero value of {\textit{P}$\mathrm{_{net}}$}$\mathrm{^{OOP}}$. 

Our study thus reveals a hitherto unknown two-step mechanism in the polarization dynamics of Aurivillius thin film that is governed by the subsequent deposition of the two building blocks of the unit cell. In the first step, the perovskite blocks covering the surface exhibit a uniform dipole orientation leading to a {\textit{P}$\mathrm{_{net}}$}$\mathrm{^{OOP}}$ build-up. In the second step, the spontaneous formation of the top polarizing Bi$_{2}$O$_{2}$ coverage induces the Aurivillius antipolar ordering in the \textit{c}-axis and hence, the suppression of {\textit{P}$\mathrm{_{net}}$}$\mathrm{^{OOP}}$.
Our polarization tracking during growth thus reveals an interfacial poling effect that is induced by the lattice chemistry. This process differs drastically from the conventional approach towards polarization engineering that relies on depolarizing-field tuning. Finally, the mechanism of atomic-scale poling should be distinguished from the spontaneous interfacial charge ordering reported in atomically sharp geometric ferroelectric$|$dielectric multilayers \cite{holtz_dimensionality-induced_2021, cheng_interface_2018}. In those, the existence of 2+/3+ mixed valence states in sub-unit-cell Fe or Mn layers generates specific charge states at the interface, which influence the polar displacement in YMnO$_{3}$ or LuFeO$_{3}$ geometric ferroelectrics. In our present work, in contrast, we reveal a chemical poling of the ferroelectric perovskite blocks constituting the Aurivillius unit cell, enabled by the deposition of chemically, structurally distinct, and highly charged Bi$_{2}$O$_{2}$ sheets.

Next, we take our lattice-chemistry-induced poling beyond the realm of the pure Aurivillius compounds and demonstrate the incorporation of complementary functional perovskite blocks into the polarizing Aurivillius framework. For this, we choose BiFeO$_{3}$ (BFO), one of the most studied perovskite ferroelectrics and the only robust room-temperature-multiferroic system \cite{fiebig_evolution_2016,catalan_physics_2009}. Furthermore, its crystal structure matches that of BFTO; see Fig. 3a for the epitaxial relationship between pseudocubic (001)$\mathrm{_{pc}}$-oriented BFO and (001)-oriented BFTO/NGO \cite{gradauskaite_robust_2020,keeney_persistence_2020}. By alternating deposition from BFO and BFTO targets (see Methods) we assemble a heterostructure with BFO embedded into the BFTO environment with its framework of Bi$_{2}$O$_{2}$ polarizing layers (Fig. 3b). We refer to these multilayers as ``composite Aurivillius'' thin films. As before, we track {\textit{P}$\mathrm{_{net}}$}$\mathrm{^{OOP}}$ in real-time by monitoring the growth by ISHG. 

The ISHG yield during the growth of a composite Aurivillius film with seven repetitions, (BFO$|$BFTO)$_{7}$ on La$_{0.7}$Sr$_{0.3}$MnO$_{3}$-buffered NGO substrate, is shown in Fig. 3c. Here, the lattice-matching La$_{0.7}$Sr$_{0.3}$MnO$_{3}$ epitaxial layer serves as a bottom electrode for ex-situ PFM characterization. We verified that its insertion does not influence the ISHG signal during the growth of the composite, as shown in Supplementary Information, Fig. S3. During the growth of a BFO layer, we observe a steep rise in the ISHG signal, indicating an increase of {\textit{P}$\mathrm{_{net}}$}$\mathrm{^{OOP}}$ \cite{de_luca_nanoscale_2017}. With the subsequent BFTO deposition, by contrast, we observe a reduction in the ISHG signal, which points to the decrease of {\textit{P}$\mathrm{_{net}}$}$\mathrm{^{OOP}}$. The growth of the polarizing Bi$_{2}$O$_{2}$ layers (as part of the BFTO) most likely forces the polarization in the composite layer to reorient. As in the case of the pure BFTO, the polarizing effect of the lattice chemistry creates an antipolar arrangement in the film, thus reducing {\textit{P}$\mathrm{_{net}}$}$\mathrm{^{OOP}}$.
This is in stark contrast with the seminal report of monitoring the evolution of the polarization during the growth of well-established ferroelectric single- or multi-layers exhibiting well-defined polarization orientations \cite{de_luca_nanoscale_2017, yu_interface_2012,efe_engineering_2024}. Here, instead, we track the reorientation of the electric dipoles in the films, induced by the capping of the charged Aurivillius-type Bi$_{2}$O$_{2}$ sheets.
Hence, the oscillatory behavior of {\textit{P}$\mathrm{_{net}}$}$\mathrm{^{OOP}}$ with alternation of the deposition target indicates the successful integration of the BFO perovskite blocks into the layered Aurivillius architecture.
The achievement of high-quality composites with a low density of structural defects is further confirmed by low magnification STEM images and X-ray diffraction data, see Supplementary Information Fig. S4. This and the ISHG signal corroborate the high level of control we achieve over the coverage of the surface with the polarizing Bi$_2$O$_2$ layers and, consequently, the electric-dipole ordering.
Most strikingly, and in contrast to the growth of single-phase BFTO (Fig. 2b), the oscillatory behavior is superimposed with an overall increase of {\textit{P}$\mathrm{_{net}}$}$\mathrm{^{OOP}}$. The completion of each composite-Aurivillius unit cell leaves a remanent net polarization that adds up unit-cell by unit-cell in {\textit{P}$\mathrm{_{net}}$}$\mathrm{^{OOP}}$, unlike with out-of-plane antipolar pure BFTO. For reference, we show the ISHG signal during the growth of the pure \textit{n}=8 Aurivillius film. It exhibits a saw-tooth-like behavior as expected for pure, even \textit{n} Aurivillius BFTO compounds, see Fig. S5. 

Let us now investigate the crystal structure associated with such polarization dynamics. The X-ray diffraction scan of our composite Aurivillius film of $\sim$30-nm thickness is shown in Fig. 3d and matches one-to-one with a pure Aurivillius \textit{n}=8 thin film, serving here as the reference. This shows that our composite Aurivillius film exhibits a perfect stacking of eight perovskite planes interleaved with fluorite-like Bi$_{2}$O$_{2}$ layers. The peak intensities and the presence of thickness fringes corroborate the excellent crystalline quality \cite{sun_progress_2021}. Reciprocal space mapping shows that the composite is fully strained to the NGO substrate, see Supplementary Information, Fig. S6. STEM imaging in Fig. 4a further confirms the high structural quality of the composite Aurivillius film.
In addition, energy-dispersive X-ray (EDX) spectroscopy elemental mapping (Fig. 4b) shows atomically sharp interfaces between the Fe- and Ti-containing perovskite blocks in the composite heterostructure with no significant intermixing, the signature of successful BFO insertion into the Aurivillius framework. 
Such ability suggests new avenues bringing the pulsed laser deposition process closer to the ultimate oxide thin film design capacity offered by molecular beam epitaxy \cite{mundy_atomically_2016,holtz_dimensionality-induced_2021}.

In order to scrutinize the consequences of the BFO insertion on the final polarization state, we use HAADF-STEM and map the atomic-scale displacements and corresponding electric-dipole distribution in our composite Aurivillius heterostructure (see the Methods section for further details). Figure 4c shows an image of the film acquired along the [$\mathrm{\bar{1}\bar{1}}$0] zone axis with the electric-dipole map plotted as arrows superimposed on the STEM image. The corresponding profiles of the \textit{B}-cation displacement $\delta$ along the [$\mathrm{\bar{1}}$10] and [001] directions are shown in Fig. 4d with $\delta$$_{x}$ and $\delta$$_{y}$ as in-plane and out-of-plane components, respectively. 
The $\delta$$_{x}$ profile clearly shows that the composite structure exhibits the same uniform in-plane polarization as the pure Aurivillius compounds. However, the out-of-plane polarization configuration differs strikingly.
Instead of the antipolar dipole ordering of even \textit{n}-numbered pure Aurivillius film (see Fig. 1b), the antipolar order of BFTO and ferroelectric order of BFO add up and as a result, we now observe a ferrielectric-like arrangement. This is consistent with the overall increase of the ISHG signal and with it of {\textit{P}$\mathrm{_{net}}$}$\mathrm{^{OOP}}$ in Fig. 3c. The asymmetry in the electric-dipole distribution here is the result of the clear separation of the Fe- and Ti-containing planes in the composite half unit cell (Fig. 4b). Our cross-sectional STEM investigation further reveals the existence of buried in-plane oriented domains within the volume of the composite films (see Supplementary Information Fig. S7.)

Finally, we now move on to the investigation of the functionality of our composite film. The films exhibit a low roughness (Fig. 5a), which enables high-resolution lateral piezoresponse force microscopy (LPFM). In the pristine state, the LPFM signal reveals the presence of oppositely in-plane-polarized domains \cite{gradauskaite_robust_2020} along the NGO [010] direction (Fig. 5b), identical to the ones of pure BFTO \textit{n}=4 thin films (Fig. S1b). Further PFM investigations carried out along two orthogonal in-plane substrate directions confirm that the uniaxial in-plane anisotropy of the pure BFTO parent compound, rather than the fourfold anisotropy of pure BFO \cite{catalan_physics_2009}, rules in the composite, see Supplementary Information Fig. S8. To verify the switchability of the out-of-plane ferrielectric-like state, we perform local poling experiments by applying a DC voltage to the scanning PFM tip. 
The vertical-PFM signal measured after out-of-plane poling of a box-shaped region strikingly confirms the switchability of ferrielectric-like {\textit{P}$\mathrm{_{net}}$}$\mathrm{^{OOP}}$ in our composite structure despite the even number of perovskite layers, associated with zero {\textit{P}$\mathrm{_{net}}$}$\mathrm{^{OOP}}$ in pure Aurivillius compounds, see Fig. 5c. This switching event is accompanied by a local {\textit{P}$\mathrm{_{net}}$}$\mathrm{^{IP}}$ reversal (Fig. 5d). The remanence of the poled pattern was verified over the time of a year, as shown in Supplementary Information Fig. S9. Furthermore, local piezoresponse switching spectroscopy revealed the desired remanent hysteretic behavior in the PFM phase in both the in-plane and, most importantly, the out-of-plane polarization direction, see Fig 5e. The amplitude curves recorded simultaneously are shown in Supplementary Information Fig. S10.

To investigate the magnetic signature of the multiferroic BFO within the composite Aurivillius structure we perform scanning nitrogen-vacancy (NV) magnetometry \cite{degen08,rondin14}. The clear magnetic contrast in Fig. 5f, which does not change as a function of the applied external magnetic field (Fig. S11a--c), 
reveals the presence of an antiferromagnetic domain configuration that is typical for a pure BFO thin film of similar thickness \cite{dufour_onset_2023}. By contrast, no magnetic response is observed from the pure Aurivillius films (Fig. S11d). This highlights the exceptional fusion of electric and magnetic functionalities achieved by integrating BFO into the Aurivillius framework. Here, the preserved ferroelectric in-plane polarization and the new ferrielectric-like out-of-plane polarization in the final composite structure coexist with an antiferromagnetic ordering. Our results demonstrate our capacity to achieve new properties while preserving the technologically relevant functionalities of the parent compounds. Finally, we put efforts into verifying the compatibility of another perovskite ferroelectric, BaTiO$_{3}$, with our Aurivillius framework. The in-situ oscillating ISHG signal during the monitoring of the growth using alternating BFTO and BaTiO$_{3}$ targets (Supplementary Information, Fig. S12) and the structural and chemical investigation of the BaTiO$_{3}$-based composite by STEM (Supplementary Information, Fig. S13) confirms the preservation of the Aurivillius-type layering. This demonstration points to the generality of our chemical poling approach and suggests new avenues in the design of functional composites based on the rich variety of properties hosted by perovskite oxide materials \cite{coll_towards_2019}.

\raggedbottom
In summary, we engineer lattice chemistry and interfacial poling effects to control the electric-dipole ordering in a ferroelectric oxide superlattice. Enabling atomic-scale tuning of the polarization direction, this approach stands out as an alternative to the conventional strategies employing depolarizing-field or charge ordering in superlattices \cite{holtz_dimensionality-induced_2021, cheng_interface_2018}. By tracking the polarization dynamics in real time, we reveal the interfacial poling effects of charged layers within Aurivillius compounds and control their stacking in a superlattice. Sandwiching a multiferroic BFO compound in between two Aurivillius Bi$_{2}$O$_{2}$ polarizing layers breaks the symmetry of the pristine Aurivillius antipolar ordering and stabilizes a ferrielectric-like ordering in the final composite structure. In this structure, the switchable net polarization coexists with antiferromagnetic order, showcasing a successful merging of the properties of the parent compounds, in-plane ferroelectric BFTO, and multiferroic BFO, while achieving additional functionalities. Furthermore, we demonstrate the generality of our approach by inserting BaTiO$_{3}$ into the Aurivillius framework and we point out the versatility of our lattice chemistry-based poling approach for the control of symmetry-driven phenomena beyond the realm of ferroelectric materials. The structural compatibility of the Aurivillius framework with many perovskite oxides exhibiting magnetic ordering or correlated-electron phenomena may even open routes for tuning magnetic textures \cite{gradauskaite_magnetoelectric_2025} or superconducting states using lattice chemistry poling effects.

\newpage

\section*{Methods}
\label{sec:Methods}

\subsection{Thin film growth and structural characterization}

The thin films and heterostructures were grown on NGO (001)$\mathrm{_O}$ substrates by pulsed laser deposition, using a KrF excimer laser at 248 nm. X-ray diffraction measurements to characterize the crystallinity, the thin-film orientation, and the epitaxial relationship between film and substrate were performed on a four-cycle thin-film diffractometer using Cu K$_{{\alpha}1}$ X-ray radiation (PanAnalytical X'Pert3 MRD).

\noindent
\textbf{Growth of pure Aurivillius BFTO \textit{n}=4 films}

The laser fluence, repetition rate, substrate temperature, and O$\mathrm{_2}$ partial pressure set for the growth of pure BFTO \textit{n}=4 films are 0.9 J cm$^{-2}$, 2 Hz, 700 {\textdegree}C, and 0.015 mbar. The thickness of the thin films was monitored using a combination of reflection high-energy electron diffraction during growth and X-ray reflectivity ex-situ.

\noindent
\textbf{Growth of composite Aurivillius film with BFO}

The laser fluence, repetition rate, substrate temperature, and O$\mathrm{_2}$ partial pressure set for the growth of the constituent layers in the composite Aurivillius film are as follows: LSMO: 0.9 J cm$^{-2}$, 1 Hz, 650 {\textdegree}C, 0.035 mbar O$_2$; first unit cell of BFTO: 0.9 J cm$^{-2}$, 2 Hz, 630 {\textdegree}C, 0.075 mbar O$_2$; BFO layer in the superlattice: 0.9 J cm$^{-2}$, 8 Hz, 640 {\textdegree}C, 0.12 mbar O$_2$; and BFTO layer in the superlattice: 0.9 J cm$^{-2}$, 2 Hz, 640 {\textdegree}C, 0.12 mbar O$_2$. To control the layer stacking of BFO and BFTO precisely, we relied solely on ISHG and stopped the BFTO growth at the minimum of the {\textit{P}$\mathrm{_{net}}$}$\mathrm{^{OOP}}$-related ISHG signal (Fig. 3c) as a sign of complete capping of the surface with the Bi$_{2}$O$_{2}$ layer. The thickness of the layers was measured by ex-situ X-ray reflectivity.

\textbf{Growth of composite Aurivillius film with BTO}

The laser fluence, repetition rate, substrate temperature, and O$\mathrm{_2}$ partial pressure set for the growth of the constituent layers in the composite Aurivillius film are as follows: LSMO: 0.9 J cm$^{-2}$, 2 Hz, 650 {\textdegree}C, 0.025 mbar O$_2$; first unit cell of BFTO: 0.97 J cm$^{-2}$, 2 Hz, 625 {\textdegree}C, 0.075 mbar O$_2$; BTO layer in the superlattice: 0.97 J cm$^{-2}$, 4 Hz, 650 {\textdegree}C, 0.015 mbar O$_2$; and BFTO layer in the superlattice: 0.97 J cm$^{-2}$, 1 Hz, 650 {\textdegree}C, 0.015 mbar O$_2$. To control the layer stacking of BTO and BFTO precisely, we relied solely on ISHG and stopped the BFTO growth at the minimum of the {\textit{P}$\mathrm{_{net}}$}$\mathrm{^{OOP}}$-related ISHG signal as a sign of complete capping of the surface with the Bi$_{2}$O$_{2}$ layer.

\subsection{Optical ISHG}
ISHG measurements were performed inside the pulsed laser deposition growth chamber in a reflection geometry under a 45{\textdegree} angle of incidence.  The probe beam of 860 nm wavelength, with a pulse duration of 60 fs, and a repetition rate of 1 kHz was generated using an amplified Ti:Sapphire laser system. The pulse energy of the probe beam was set to 20 $\mathrm{\mu}$J. The beam was focused onto the sample with a probe diameter of 250 $\mathrm{\mu}$m. The generated SHG signal at 430 nm was selected using a monochromator and converted into an electrical signal with a photomultiplier tube and gate electronics \cite{de_luca_nanoscale_2017}. To capture {\textit{P}$\mathrm{_{net}}$}$\mathrm{^{OOP}}$ and {\textit{P}$\mathrm{_{net}}$}$\mathrm{^{IP}}$ separately, the linear polarization of the incident probe beam and of the frequency-doubled SHG light were fixed to 90{\textdegree} (p-polarized) and 0{\textdegree} (s-polarized), respectively \cite{de_luca_nanoscale_2017}.

\subsection{Scanning transmission electron microscopy}
Electron-transparent samples for STEM were prepared for cross-section imaging by using an FEI Helios 660 G3 UC focused gallium (Ga$^{+}$) ion beam (FIB) operated at acceleration voltages of 30 and 5 kV after deposition of C and Pt protective layers. STEM was performed on an analytical FEI Titan Themis, operated at 300 kV and equipped with a spherical-aberration probe corrector (CEOS DCOR) and Super-X EDX technology. The experiments were performed by setting the beam semi-convergence angle to 18 mrad in combination with an annular semi-detection range of the HAADF detector of 72--200 mrad. The EDX elemental maps were recorded with a beam current of 290 pA and a dwell time of 10 $\mu$s/pixel.

\noindent
\textbf{Image processing and data analysis}

The HAADF-STEM images used for the polarization mapping were obtained as average signals of time-series of 12 frames (2048 $\times$ 2048 pixels at a dwell time of 1 $\mu$s), after rigid and nonrigid registrations performed using the Smart Align software \cite{jones_smart_2015}. The obtained images were then filtered in Fourier space using a bandpass filter and the atomic column fitting was performed using the Python library Atomap \cite{nord_atomap_2017} after applying a Python-based probe deconvolution algorithm on the HAADF-STEM images. 

In perovskite ferroelectrics with the general formula \textit{AB}O$_3$, the polar displacement resulting from the off-centering of the cations with respect to the oxygen octahedra induces spontaneous polarization. Thus, polarization maps can be efficiently calculated from the relative displacements of the two cationic sublattices as the \textit{B}-cations move along with the oxygen octahedra. Here, we determine the displacement vector by measuring the polar displacement of the B position in the image plane from the center of its four \textit{A}-neighbors. Finally, in the polarization maps the polarization vectors are plotted opposite to the polar displacement of the \textit{B}-cations. 

The EDX maps for Bi, Ti, and Fe were obtained as time-averages of 130 frames, by integrating the Bi--M, Ti--K, and Fe--K edges intensities, respectively.

\subsection{Ex-situ Atomic force microscopy and PFM measurements}
The surface topography, the ferroelectric domain configuration, and the local switching characteristics of the pure BFTO \textit{n}=4 and composite Aurivillius thin films were studied using a Bruker Multimode 8 atomic force microscope equipped with Pt-coated Si tips (MikroMasch, \textit{k}= 5.4 N m$^{-1}$).

Our LPFM and VPFM images are recorded simultaneously in Cartesian coordinates (using X and Y outputs of the lock-in amplifiers, rather than R and $\theta$ for amplitude and phase). In this way, we do not lose any polarization information and minimize instrumental background piezoresponse interfering with the measurements. The procedure for our PFM measurement calibration is detailed elsewhere \cite{jungk_quantitative_2006,jungk_consequences_2007,gradauskaite_robust_2020}. The PFM measurements were performed using a drive voltage of V$\mathrm{_{ac}}$ = 1.5 V for pure BFTO \textit{n}= film and V$\mathrm{_{ac}}$ = 5 V for composite Aurivillius thin film at 10 kHz. The poling of the composite Aurivillius film was performed by applying a bias voltage V$\mathrm{_{dc}}$= $-$10 V. 

\subsection{Local piezoresponse switching spectroscopy}

Local piezoresponse switching spectroscopy was performed on an Asylum Research Cypher S HV atomic force microscope using a platinum-coated AFM tip with a nominal resonance frequency of 75 kHz and spring constant of 2 N m$^{-1}$. Following the standard switching spectroscopy PFM protocol \cite{jesse_switching_2006}, a triangular comb DC bias of up to 20 V was applied simultaneously with a dual 1 V AC-excitation voltage to the tip at the contact resonance frequency of 200--300 kHz for vertical and 800--1000 kHz for lateral response characterization. The response was acquired and processed with the integrated software from Asylum Research, yielding hysteretic voltage-dependent amplitude and phase response.

\subsection{Scanning NV magnetometry}

Magnetometry experiments were performed at ambient conditions with a custom-built scanning NV microscope (details in \cite{huxter23}). Spatially resolved stray magnetic field images from the BFO/BFTO composite Aurivillius sample were acquired by monitoring the spin-dependent photoluminescence of a raster-scanned NV center located inside a (110)-cut diamond probe (purchased from QZabre AG). Specifically, a spectrum demodulation technique \cite{welter23} was used for the magnetometry. A small external bias field ($\sim$ 5 mT) was applied along the NV symmetry axis (described by the polar and azimuthal angles of $\mathrm{\theta}$ $\sim$38{\textdegree} and $\mathrm{\phi}$ $\sim$85{\textdegree}, respectively) to separate the degeneracy of the ms = $\pm$1 spin levels \cite{rondin14}. Sweeping the frequency of a microwave source ($\sim$3 GHz) through one of the NV center's spin resonances at kilohertz frequencies, in combination with continuous laser excitation, produced a periodic photoluminescence signal whose phase encodes the local magnetic field projection along the NV symmetry axis. Imaging speeds were 4--16 pixels per second.

\section*{Acknowledgements}
\label{sec:Acknowledgements}

I.E. and M.T. acknowledge the Swiss National Science Foundation under project no. 200021{\_}188414. I.E. and M.T. thank Daniele Leonelli, Rui Chen, Wei Li, and Jeffrey Xu for fruitful discussions. A.V. and M.D.R. acknowledge support by the Swiss National Science Foundation (SNSF) under project no. 200021{\_}175926. W.S.H. acknowledges funding from the European Research Council through ERC CoG 817720 (IMAGINE). C.L.D. acknowledges funding from the SNSF, under Grant. No. 200020{\_}212051 and from the SBFI, Project ``QMetMuFuSP" under Grant. No. UeM019-8, 215927.

\section*{Data Availability}
\label{sec:Data Availability}
The data support the findings of this study are available from the corresponding authors upon request.

\begin{figure}[p]
    \centering
    \includegraphics[scale=0.50, trim={1cm 0cm 1cm 0cm}, clip]{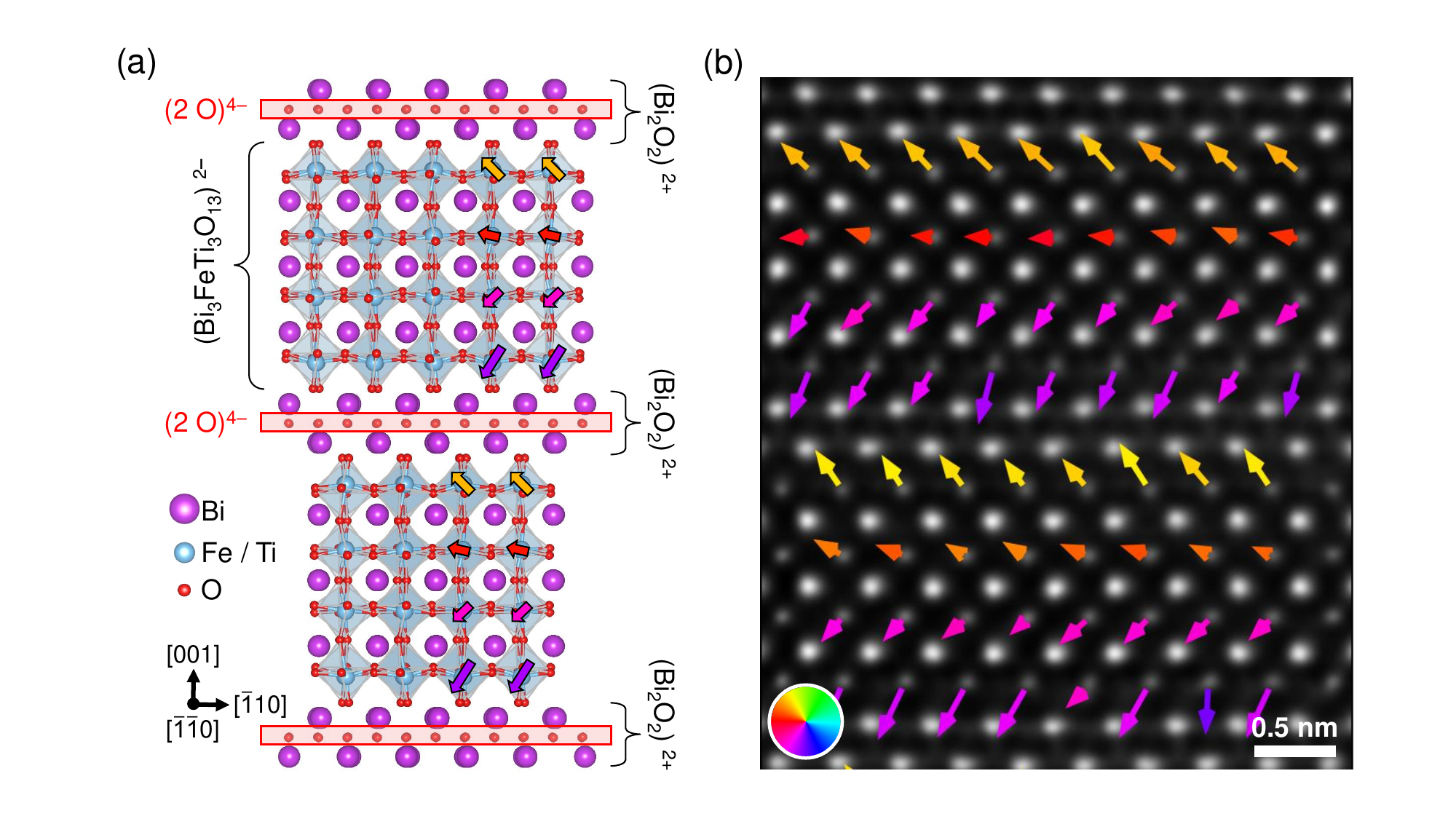}
    \caption{\textbf{Antipolar ordering of electric dipoles in c-oriented BFTO.} a) Schematic of the BFTO unit cell with its two constituent units: four perovskite layers and Bi$_{2}$O$_{2}$ sheets. The negatively charged oxygen atomic layer in the Bi$_{2}$O$_{2}$ is highlighted in red. The arrows represent the electric dipoles pointing toward the nearest Bi$_2$O$_2$ layer. b) HAADF-STEM image of the BFTO unit cell with the measured electric dipole distribution overlaid. Visualization of the BFTO sample along the BFTO [110] direction was deliberately chosen to study the polarization rotation across domain walls and structural defects, as detailed in reference \cite{campanini_buried_2019}. The arrows show the polarization vector at each \textit{B}-site cation position, the arrows are colored according to the given 360{\textdegree} color wheel. The image without the overlaid arrows is presented in the Supplementary Information, Fig. S14a.}
    \label{fig:fig1}
\end{figure}

\begin{figure}[p]
    \centering
    \includegraphics[scale=0.5, trim={0cm 9cm 0cm 9cm}, clip]{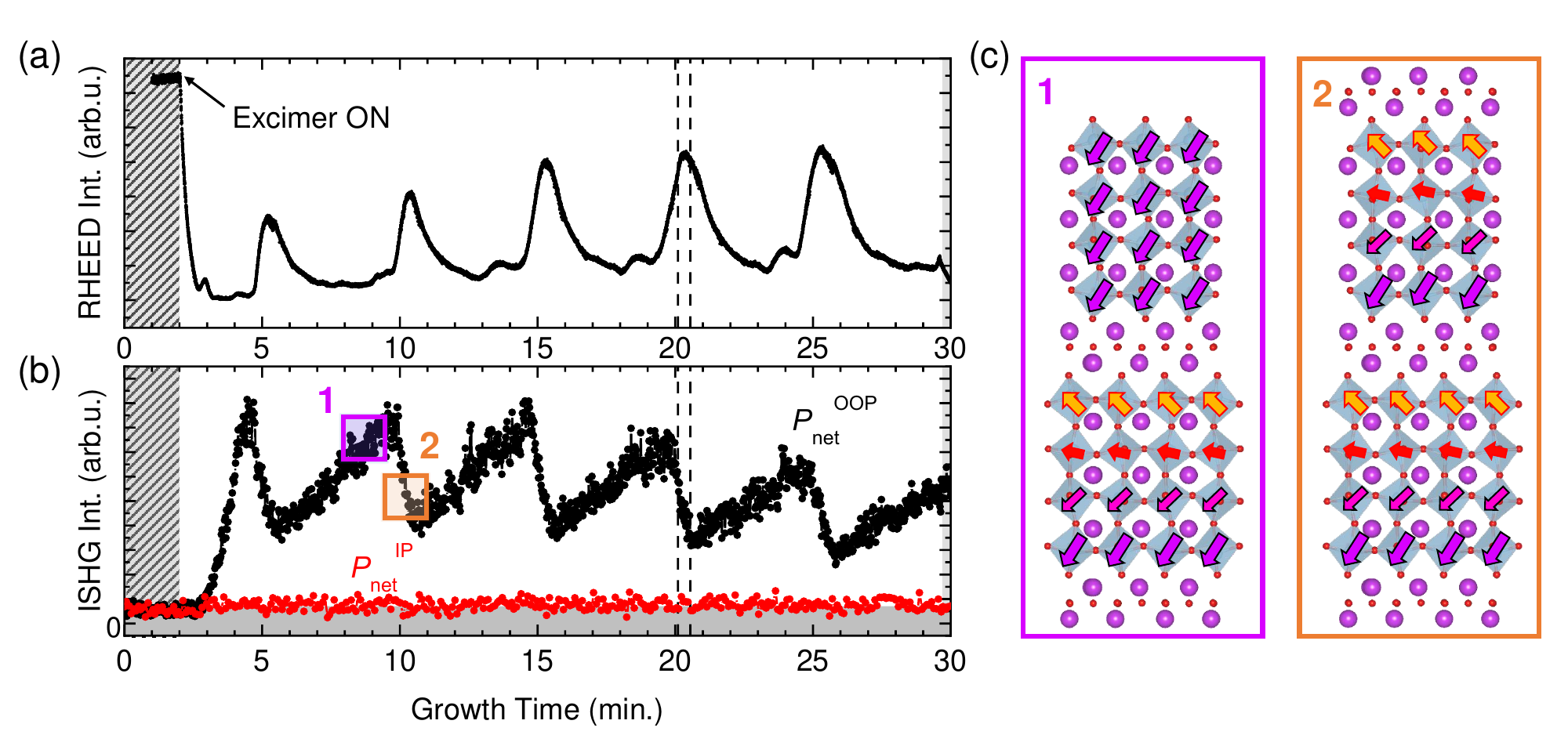}
    \caption{\textbf{Real-time monitoring of the polarization dynamics of a pure BFTO \textit{n} $=4$ film.} Simultaneously recorded a) RHEED and b) ISHG signals during the growth of the BFTO films. The oscillating RHEED intensity in (a) ensures a 2D layer-by-layer growth mode of BFTO thin film. In (b) the {\textit{P}$\mathrm{_{net}}$}$\mathrm{^{IP}}$-related ISHG signal (red) stays constant at the paraelectric background (dark gray) level, while the {\textit{P}$\mathrm{_{net}}$}$\mathrm{^{OOP}}$-related ISHG signal (black) oscillates in a saw-tooth-like manner. Dashed lines highlight the out-of-phase relationship between the RHEED and ISHG oscillations. The signal level never drops to the paraelectric background level because the interface created between the substrate and thin film contributes to the signal. c) Schematics of the electric-dipole orientation within the BFTO unit-cell (1) at the maximum and (2) during the sharp drop of the {\textit{P}$\mathrm{_{net}}$}$\mathrm{^{OOP}}$-related ISHG signal. 
    \label{fig:fig2}}
\end{figure}

\begin{figure}[p]
    \centering
    \includegraphics[scale=0.5, trim={1cm 0cm 1cm 0cm}, clip]{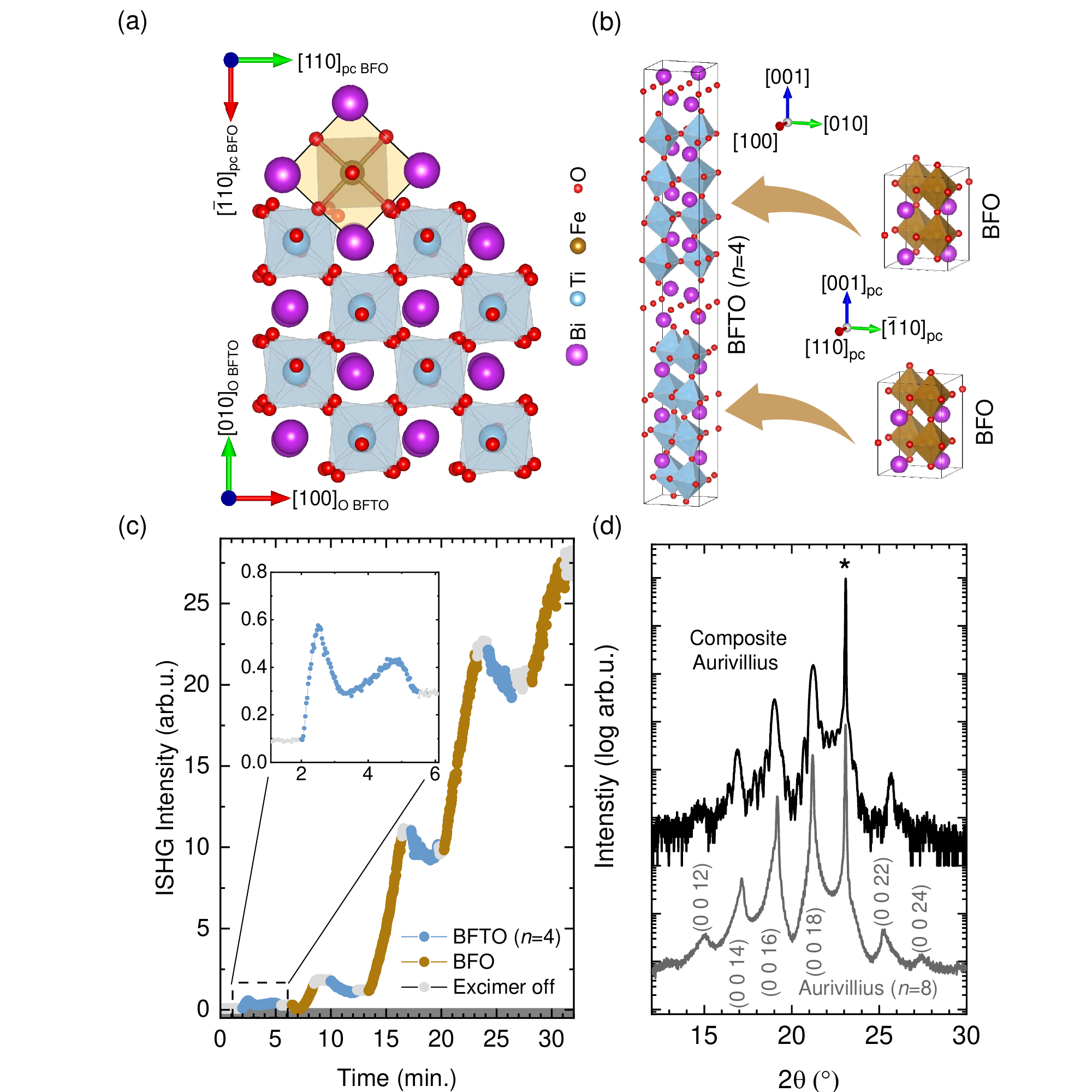}
    \caption{\textbf{Integration of BFO into the Aurivillius framework.} a) Epitaxial relationship between BFTO and BFO in the composite. The Aurivillius [100] polarization direction aligns with the orthorhombic [010] direction of the NGO substrate. b) Schematic of our lattice chemistry engineering approach that we apply for the growth of composite Aurivillius thin film. c) {\textit{P}$\mathrm{_{net}}$}$\mathrm{^{OOP}}$-related ISHG signal recorded during the growth of composite Aurivillius thin films. The paraelectric background level is highlighted in dark gray. d) {$\theta$}-2$\theta$ X-Ray diffraction patterns of the composite Aurivillius (black) and a pure Aurivillius \textit{n}=8 film (gray) around the NdGaO$_{3}$ (002)$\mathrm{_{O}}$ substrate peak (asterisk). The small peak shift in 2$\theta$ is most likely caused by the cation ordering in the composite structure and a pronounced reduction of the tetragonality of Fe-rich perovskite blocks. 
    \label{fig:fig3}}
\end{figure}

\begin{figure}[p]
    \centering
    \includegraphics[scale=0.5, trim={0cm 0cm 0cm 0cm}, clip]{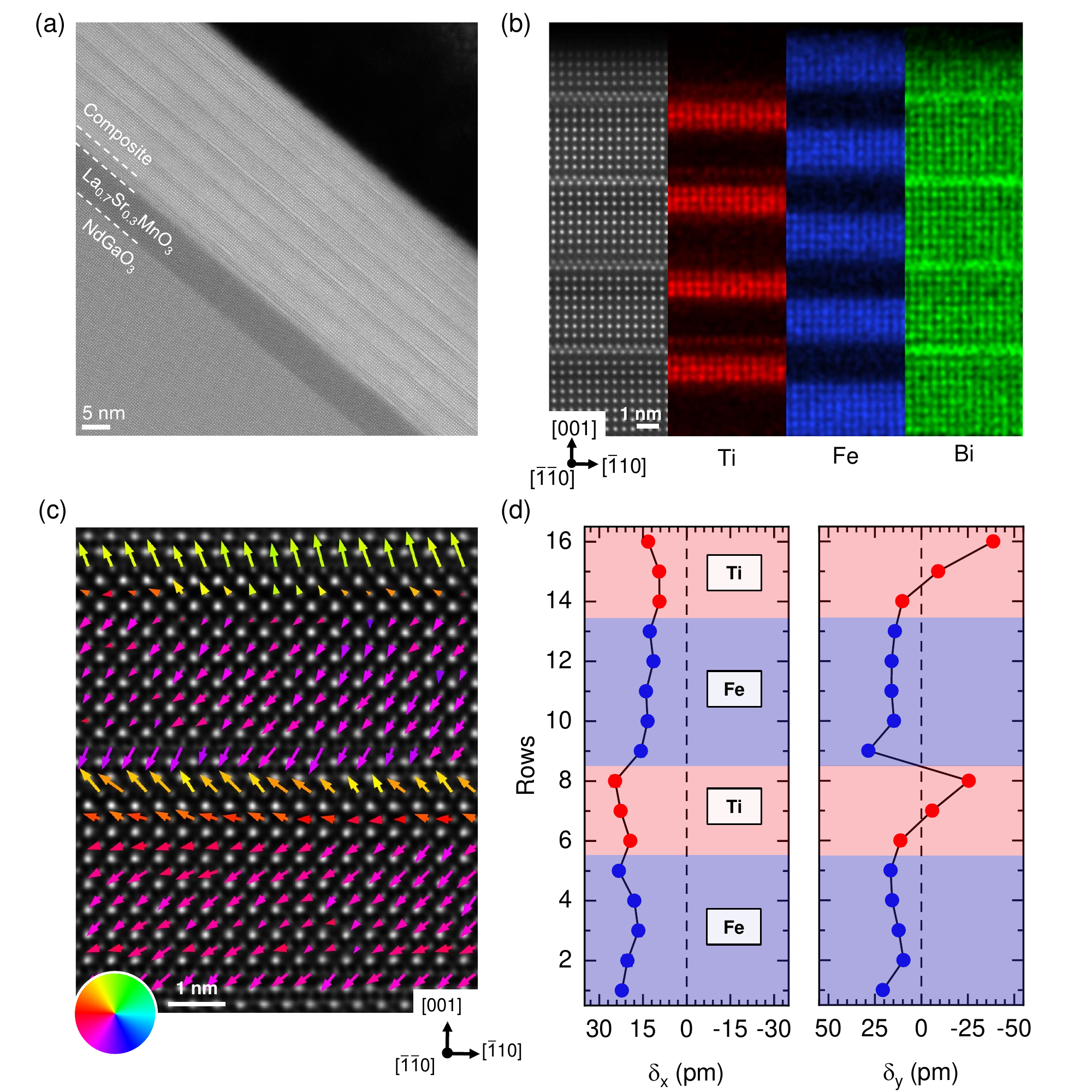}
    \caption{\textbf{Atomically sharp interfaces in the lattice-chemistry-engineered BFO/BFTO composite film and the resulting ferrielectric-like order} Low-magnification HAADF-STEM image of the BFO/BFTO composite on the La$_{0.7}$Sr$_{0.3}$MnO$_{3}$-buffered NGO substrate. Note that the BFTO \textit{n} $=4$ template lies between the La$_{0.7}$Sr$_{0.3}$MnO$_{3}$ buffer and the composite. The low density of structural defects corroborates the high structural quality of the composite film inferred by the ISHG and XRD. b) High-magnification HAADF-STEM image and corresponding EDX-mapping of the composite Aurivillius thin film, showing atomically sharp interfaces of internal layering of the Ti- and Fe-cation distributions. Note that the pronounced Fe ordering near the lower Bi$_2$O$_2$ layer renders the interface between the Fe-containing perovskite blocks originating from the BFTO and BFO targets indistinguishable. Here, we estimate the composite to contain five BFO-unit-cells. c) The HAADF-STEM image is overlaid with arrows representing the orientation of the electric dipoles. The arrows show the polarization vector at each \textit{B}-site cation position, the arrows are colored according to the given 360{\textdegree} color wheel. Alternating, uneven distribution of upward- and downward-oriented electric dipoles reveals a ferrielectric-like ordering within the composite Aurivillius thin film. The image without the overlaid arrows is presented in the Supplementary Information, Fig. S14b. d) Corresponding \textit{B}-cation displacement profiles along in-plane ($\delta_{x}$, left) and out-of-plane ($\delta_{y}$, right) axes. The internal layering of the Ti-and Fe-cation distributions is shown in red and blue, respectively. 
    \label{fig:fig4}}
\end{figure}

\begin{figure}[p]
    \centering
    \includegraphics[scale=0.5, trim={5cm 4cm 5cm 4cm}, clip]{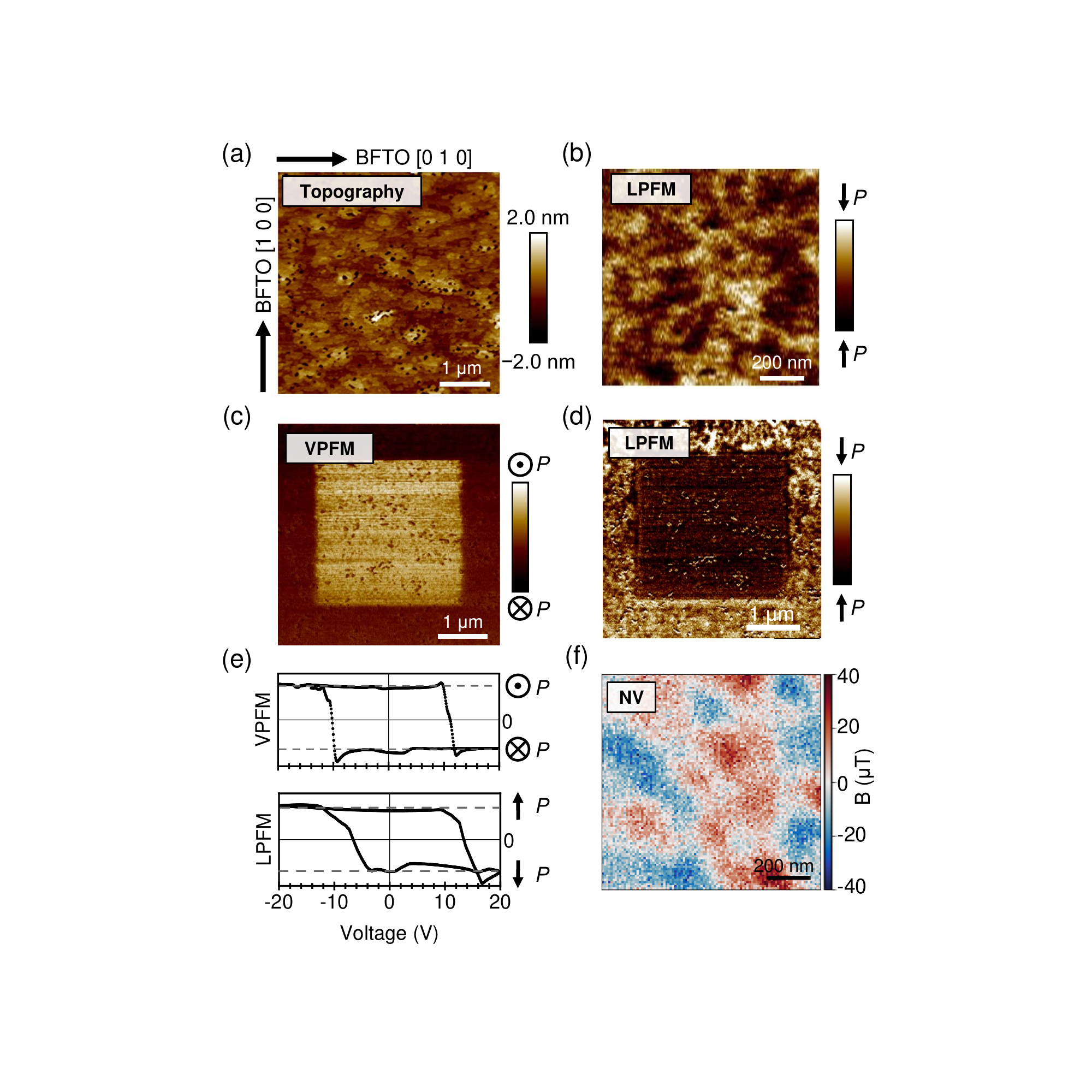}
    \caption{\textbf{Coexisting ferrielectric-like and antiferromagnetic order in the BFO/BFTO composite.} a) Atomic force microscopy scan of the surface topography of the composite Aurivillius thin film. b) Lateral-PFM (LPFM) contrast displaying the uniaxial in-plane domains, identical to the pure BFTO thin films. c) Vertical-PFM (VPFM) contrast displaying the out-of-plane component of the ferrielectric-like dipole ordering in the composite Aurivillius thin film after poling with a biased PFM tip with $-$10 V. The outer frame with darker color corresponds to the pristine {\textit{P}$\mathrm{_{net}}$}$\mathrm{^{OOP}}$-downward polarized film, while the brighter contrast in the square-shaped poled region shows {\textit{P}$\mathrm{_{net}}$}$\mathrm{^{OOP}}$-upward polarization. d) LPFM image of the box-shaped region poled by applying a $-$10 V DC-voltage to the scanning PFM tip. The uniform polarization direction in the poled region indicates the complete switching of the in-plane polarization. e) Local piezoresponse switching spectroscopy of the composite film. The phase of the VPFM and LPFM signal recorded during the poling of the film reveals a hysteretic behavior along both the in-plane and out-of-plane directions. See Supplementary Fig. S10. for the amplitude signals. f) Scanning NV magnetometry image revealing antiferromagnetic domains.       
    \label{fig:fig5}}
\end{figure}

\newpage
\bibliography{main.bib}

\end{document}